\documentclass[12pt]{article}

\catcode`\@=11

\global\arraycolsep=2pt
\usepackage{amsbsy,amssymb,latexsym,amsfonts,amsmath}
\usepackage{graphicx,color}

\allowdisplaybreaks

\begin{document}

\vspace*{0.7cm}

\begin{center}
{ \Large Further swampland constraint on Dirac Neutrino}
\vspace*{1.5cm}\\
{Taiga Harada and Yu Nakayama}
\end{center}
\vspace*{1.0cm}
\begin{center}

Department of Physics, Rikkyo University, Toshima, Tokyo 171-8501, Japan

\vspace{3.8cm}
\end{center}

\begin{abstract}
Recent studies on swampland conjectures (e.g. non-SUSY AdS conjecture or AdS distance conjecture) predict that the (lightest) neutrino must be Dirac and the mass must be cosmologically small $m < c \Lambda^{1/4}_{\text{cc}}\sim 10 \ \text{meV}$. The Dirac neutrino naturally accompanies a $U(1)$ symmetry that can be embedded in the anomaly-free $U(1)_{B-L}$ global symmetry of the standard model. We point out that the swampland conjectures applied to  a circle compactification with the $U(1)$ symmetry twisting lead to further constraints. In particular, the $U(1)$ symmetry must be broken down to $\mathbb{Z}_4$,$\mathbb{Z}_8$ or $\mathbb{Z}_{10}$ in the case of normal hierarchy, and $\mathbb{Z}_4$ in the case of inverted hierarchy, providing evidence for the absence of continuous global symmetry in quantum gravity. We also predict a more constrained upper bound  of the neutrino mass for each case.

\end{abstract}

\thispagestyle{empty} 

\setcounter{page}{0}

\newpage

\section{Introduction}
A naive application of the philosophy of the renormalization group suggests that quantum gravity has little predictive power on low energy physics. If this were the case, it would be extremely difficult to obtain any experimental clues on quantum gravity.
On the other hand, there has been growing evidence that the consistency of the quantum gravity such as the string theory does imply non-trivial constraints on the low energy effective field theories. 
Such constraints are known as swampland conjectures (see e.g. \cite{Palti:2019pca}\cite{Grana:2021zvf} for reviews).

The standard model of particle physics coupled with the Einstein gravity not only describes almost everything in our observable universe but also allows many other solutions that are not realized in our observable universe within its validity \cite{Arkani-Hamed:2007ryu}. In particular, it allows a $1+2$ dimensional compactified universe (rather than $1+3$ dimensional universe of ours) and they are subject to the swampland conjectures \cite{Heidenreich:2015nta}\cite{Montero:2016tif}\cite{Hamada:2017yji}\cite{Rudelius:2021oaz}.

The idea to study the circle compactified solution of the standard model of particle physics coupled with the Einstein gravity was pursued in the  literature in the context of various swampland conjectures. They may explain e.g. strength of the gauge coupling constant, supersymmetry breaking, electroweak hierarchy, cosmological constant or neutrino masses. See e.g. \cite{Grana:2021zvf} and reference therein.

In \cite{Ibanez:2017kvh}\cite{Gonzalo:2018dxi}\cite{Gonzalo:2021fma}\cite{Gonzalo:2021zsp}, they showed that certain swampland conjectures (i.e. non-SUSY AdS conjecture or AdS distance conjecture) predict that the (lightest) neutrino must be Dirac and the mass must be cosmologically small $m < c \Lambda_{\text{cc}}^{1/4} \sim 10 \ \text{meV}$, magically relating the cosmological constant $\Lambda_{\text{cc}}$ of the universe and the mass of the neutrino.
In this paper, we study further constraints on the property of the Dirac neutrino.
The Dirac neutrino naturally accompanies a $U(1)$ symmetry that can be embedded in the anomaly-free $U(1)_{B-L}$ global symmetry of the standard model. We will show that the swampland conjectures applied to the circle compactification with the $U(1)$ symmetry twisting lead to further nontrivial constraints on the neutrino mass as well as the fate of the $U(1)$ symmetry. Indeed, we show that the $U(1)$ symmetry must be broken down to  the discrete subgroup $\mathbb{Z}_4$, $\mathbb{Z}_8$ or $\mathbb{Z}_{10}$, providing evidence for the absence of continuous global symmetry in quantum gravity.

The rest of the paper is organized as follows. Section 2 is a review of the relevant swampland conjectures and the circle compactified standard model of particle physics. In section 3, we show our original discussions on the constraint on the Dirac neutrino from the circle compactification with the $U(1)$ symmetry twisting. In section 4, we discuss future directions to be pursued.

\section{Swampland conjectures and standard model on a circle}

Among various swampland conjectures, in this paper we focus on particular two conjectures, non-SUSY AdS conjecture \cite{Ooguri:2016pdq}\cite{Freivogel:2016qwc} and the AdS distance conjecture \cite{Lust:2019zwm}. Let us first summarize the claim of the conjectures and the underlying motivations.

The non-SUSY AdS conjecture claims ``non-supersymmetric AdS vacuum must be  unstable". This is motivated by the weak gravity conjecture: the gravitational force must be weak so that the extremal black hole can (or must) decay \cite{Arkani-Hamed:2006emk}\cite{Ooguri:2006in}. Indeed, the stronger version of the weak gravity conjecture states that the non-supersymmetric extremal black hole cannot exist. The near horizon limit of the extremal black hole gives the AdS vacuum, so the weak gravity conjecture implies that the AdS solution from the extremal black hole must be unstable. In this sense, the non-SUSY AdS conjecture is in part supported by the weak gravity conjecture, but also generalizes it because it states that no other construction is possible.

The AdS distance conjecture claims ``if we have a series of effective field theories that have adjustable vacuum energy, when it approaches the Minkowski limit from the AdS side, infinite numbers of light degrees of freedom should appear" \cite{Ooguri:2006in}\cite{Ooguri:2018wrx}\cite{Hamada:2021yxy}. This is motivated by the swampland distance conjecture: if  we move toward the infinite distance in the moduli space of effective (gravitational) theories, we must encounter an infinite tower of light degrees of freedom. The AdS distance conjecture further regards the vacuum energy as a ``moduli" and the Minkowski limit as the infinite distance (in theory space).

The philosophical foundation of our discussion below is that we are going to assume that every solution of the effective field theory within its validity is subject to these swampland constraints. Our main target is the standard model of particle physics coupled with the Einstein gravity. The validity of the swampland conjecture means that not only our universe must be consistent with the swampland conjectures, but the constraint should also apply to the other solutions. 

Consider the standard model of particle physics coupled with the Einstein gravity (with the cosmological constant). We are looking for a solution with a circle compactification  of radius $R$. Since we are interested in the regime $R> \Lambda_{\mathrm{QCD}}^{-1}$, the effective degrees of freedom that we keep track of are graviton, photon, (light) neutrinos, light leptons (i.e. electron and muon) as well as some mesons ($\pi$, $K$ and $\eta$ are included in our computation). As we will see in a moment, the number of light fermions whose mass is  of the cosmological scale will be important to find non-trivial three-dimensional AdS solutions.

\begin{table}[hbtp]
\centering
\begin{tabular}{l|l|l}
particle & mass              & DOF   \\ \hline
photon   & 0                 & 2       \\
graviton & 0                 & 2       \\
$\nu$    & $< 0.1$ eV & 6 (Majorana) or 12 (Dirac) \\
$e$      & 0.511 MeV           & 4       \\
$\mu$    & 100 MeV           & 4       \\
$\pi$    & 140 MeV           & 3       \\
$K$    & 500 MeV           & 4  \\
$\eta$    & 550 MeV           & 1       
\end{tabular}
 \caption{A list of light particles in the standard model.}
 \label{particle}
\end{table}

One of the most important assumptions hereafter is that we have no other light (yet-to-be-observed) degrees of freedom such as axion or dark radiation. The existence of extra light degrees of freedom will change the following story, which would be an interesting future direction to be studied.

Let us recall some experimental facts about the neutrino. After the first experimental evidence for the neutrino oscillation in 1998, it is now accepted that neutrino has a non-zero mass, but the nature of the mass, e.g. if they are Dirac type or Majorana type is unknown.\footnote{Of course, we could have both mass terms simultaneously if the symmetry allows; what we mean by ``Dirac" here is that the left-handed neutrino and the right-handed neutrino are (almost) degenerated.} The neutrino oscillation only gives information of the mass difference \cite{GAMBITCosmologyWorkgroup:2020rmf}:
\begin{align}
\Delta m_{12}^2 = m_2^2-m_1^2 &= 7.53(18) \times 10^{-5} \ \mathrm{eV}^2 \cr
\Delta m_{32}^2 = m_3^2-m_2^2 &= 2.44(6) \times 10^{-3} \ \mathrm{eV}^2  \ (\mathrm{NH}) \cr
\Delta m_{23}^2 = m_2^2-m_3^2 &= 2.51(6) \times 10^{-3} \ \mathrm{eV}^2 \ (\mathrm{IH})
\end{align}
It is called the normal hierarchy (NH) when $m_1 < m_2 <m_3$ and inverted hierarchy (IH) when $m_3 < m_1 <m_2$.  

We compactify the standard model of particle physics coupled with the Einstein gravity on a circle whose physical radius is $R$. The radius will be fixed by minimizing the effective potential for the radion field $R$. The potential depends on the boundary condition of various fields around the circle and all the possible boundary conditions will be subject to the swampland conjectures.

The one-loop effective potential for the radion is given by the general formula \cite{Arkani-Hamed:2007ryu}\cite{Hamada:2017yji}\footnote{We find it a little awkward that the phase is $2\pi \theta$ rather than $\theta$, but we keep following the convention used in \cite{Hamada:2017yji}.}
\begin{align}
V(R) = \frac{2\pi \Lambda_{\text{cc}}}{R^2} - \sum_p(-1)^{2s_p} n_p \frac{m_p^2}{4\pi^3 R^4} \sum_{n=1}^{\infty} \frac{K_2(2\pi  n m_p R)}{n^2} \cos (2\pi n \theta_p) \ ,
\end{align}
where $\Lambda_{\text{cc}}$ is the four-dimensional cosmological constant, $n_p$ is the number of the four-dimensional particle $p$ counted by the real degree of freedom, which has the spin $s_p$ and the mass $m_p$. We have included the twisting parameter $\theta_p$ so that the field associated with the particle $p$ is twisted by the phase $e^{2\pi i \theta_p} $ when it goes around the circle. 

The case when all the fields satisfy the periodic boundary condition with $\theta_p = 0$ was extensively studied in \cite{Gonzalo:2021fma}\cite{Gonzalo:2021zsp}. The schematic picture is that at large $R$, the net number of fermions minus bosons, whose mass scale is around the cosmological constant, determines the shape of the potential. When the net number is negative (i.e. when the lightest neutrino is Majorana), we find the (stable) AdS vacuum (under the assumption that near $R=0$, potential becomes positive, which is the case when the number of not only light but {\it total} fermions is  positive) while when the net number is positive, we find no AdS vacuum. 
In order to avoid the three-dimensional AdS vacuum, it was found that the lightest neutrino must be Dirac and its mass has an upper bound determined by the cosmological constant (i.e. $\Lambda_{\text{cc}} = 3.25 \times 10^{-11} \ \mathrm{eV}^4$ \cite{ParticleDataGroup:2006fqo}). More precisely
\begin{align}
\mathrm{NH}: m_1 \lesssim 8.3 \ \mathrm{meV} \cr
\mathrm{IH}: m_3 \lesssim 2.8 \ \mathrm{meV} \label{prediction1}
\end{align}
The difference comes because in the IH, the second and the third lightest fermions are heavier than the case of NH, and the effects to the potential at large $R$ are suppressed.

Almost the same condition was obtained by studying the AdS distance conjecture. On one side, the AdS distance conjecture appears more attractive because it is less UV sensitive than the non-SUSY AdS conjecture; the latter may be affected by the possibility or impossibility of the compactification with the UV cut-off scale. On the other hand, the concept of the ``series of  effective field theories" is not a priori known in the low energy limit, and we need to make some assumptions about the reasonable ``series".

In \cite{Gonzalo:2021fma}\cite{Gonzalo:2021zsp}, the scan was made over the neutrino mass with or without fixing the cosmological constant. For example, the homogeneous scanning
\begin{align}
m_1(\Lambda) &= \lambda m_1^{\text{exp}} \cr
m_2(\lambda) &= \lambda \sqrt{(m_1^{\text{exp}})^2 + \Delta {m}^2_{21}} \cr
m_3(\lambda) &= \lambda \sqrt{(m_1^{\text{exp}})^2 + \Delta {m}^2_{21} + \Delta m_{32}^2 }  \label{homoscan}
\end{align}
with $10^{-4}<\lambda<1$ (see \cite{Gonzalo:2018dxi} for the necessity of the lower bound) while fixing the cosmological constant leads to the same constraint alluded above \eqref{prediction1}. 
 A simple reason why we obtain the same bound is that when the non-SUSY AdS conjecture is violated with one $m_1^{\text{exp}}$, by varying $\lambda$ we will be able to obtain the Minkowski vacuum without any tower of massless states, hence violating the AdS distance conjecture as well.

\section{Constraints from twisted boundary conditions}
We now consider the swampland constraint by further varying the boundary conditions for fermions. The effect of the boundary condition induced by the gauge field was studied in \cite{Hamada:2017yji}. Our focus is the global symmetry that acts on the neutrino, which plays the most important role in our applications of the swampland conjectures.

We have already seen that by studying the swampland conjectures with $\theta = 0$, the (lightest) neutrino must be Dirac. The Dirac neutrino is naturally equipped with the $U(1)$ symmetry that can be identified with the anomaly-free $U(1)_{B-L}$ symmetry of the standard model. In order to change the boundary condition consistently, the symmetry that we use in the twisting must be anomaly-free, and the $U(1)_{B-L}$ is the most natural candidate. In our analysis, twisting by the $U(1)_{B-L}$ effectively means that all the fermions included in the potential , which appeared in table \ref{particle}, are assigned with the same $\theta$.

\begin{figure}[tb]
    \centering
    \includegraphics[keepaspectratio, scale=1.0]{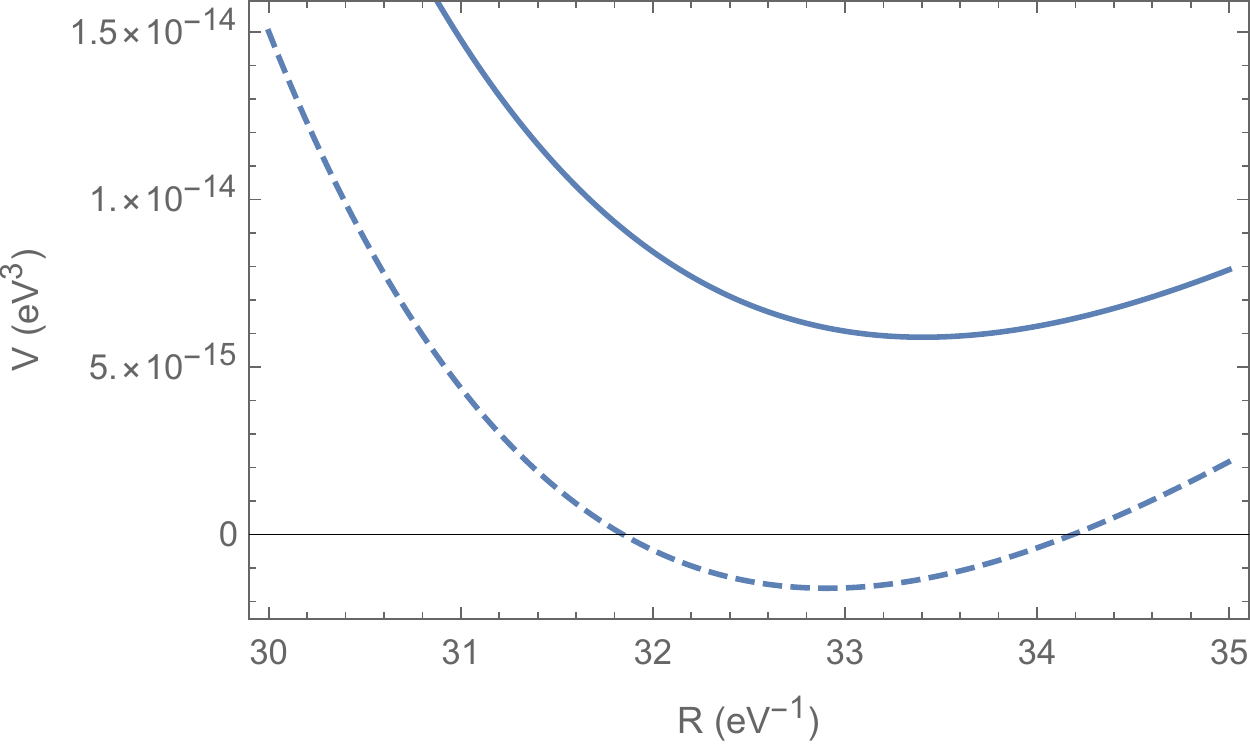}
    \caption{Radion potential at $\theta = 0.125$ (NH).  The dashed line is at $m_1 = 0.46 \ \text{meV}$. The solid line is at $m_1 = 0.45 \text{meV}$. The non-SUSY AdS vacuum appears for larger $m_1$.}
    \label{fig:hogehoge}
\end{figure}

Let us first focus on the case with NH. When $\theta \lesssim 0.13$, by adjusting the mass of the lightest neutrino, our numerical analysis shows that we can assure that the compactified vacuum  has $V > 0$, not being AdS. See Fig \ref{fig:hogehoge} as an illustration. However, when $\theta$ is larger than $0.131 (>\frac{1}{8})$, the vacuum becomes AdS no matter how we adjust the mass of the lightest neutrino, which means that it violates the non-SUSY AdS conjecture.

The non-SUSY AdS conjecture remains violated up to $\theta \sim 0.197(1) (<\frac{1}{5})$, above which the would-be AdS vacuum rolls down to $R=0$ so that the violation of the non-SUSY AdS conjecture could be avoided. The details will depend on the situation near $R=0$, and the neutrino mass alone cannot say anything about it.\footnote{Indeed, the number $0.197(1)$ and whether it allows $\mathbb{Z}_{10}$ (see below) is UV sensitive. Our computation here includes $K$ and $\eta_8$, but the cut-off energy seems arbitrary.} This continues to be the case  up to $\theta = \frac{1}{2}$.

The simple picture to explain the above numerical analysis is when $\theta$ varies from $0$ to $\frac{1}{2}$, the contributions of fermions to the vacuum energy becomes more like those of bosons, so the net effective number of light fermion decreases when we increase $\theta$. When $\theta$ becomes much larger, then the potential becomes run-away toward $R=0$ because the positivity of the potential near $R=0$ due to the excess of total (effective) number of fermions is now lost. In this way, we see that a certain intermediate range of $\theta$ is in the swampland.

\begin{figure}[tb]
    \centering
    \includegraphics[keepaspectratio, scale=0.9]{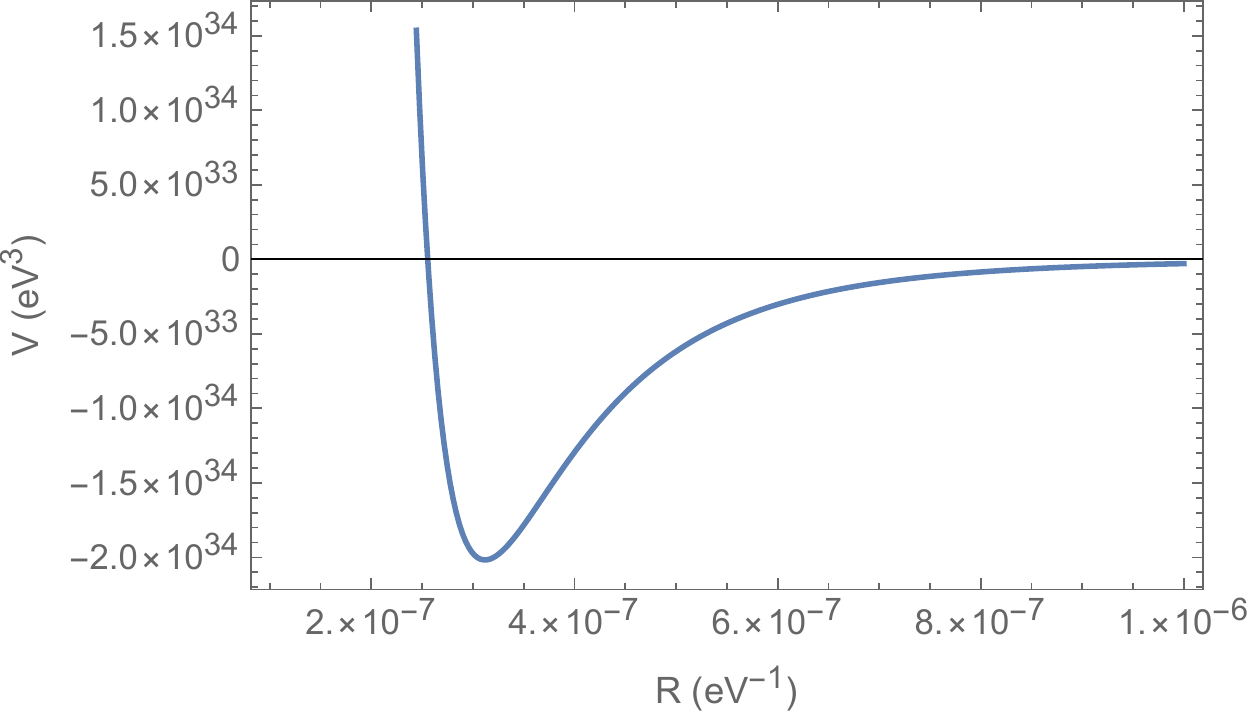}
        \includegraphics[keepaspectratio, scale=0.9]{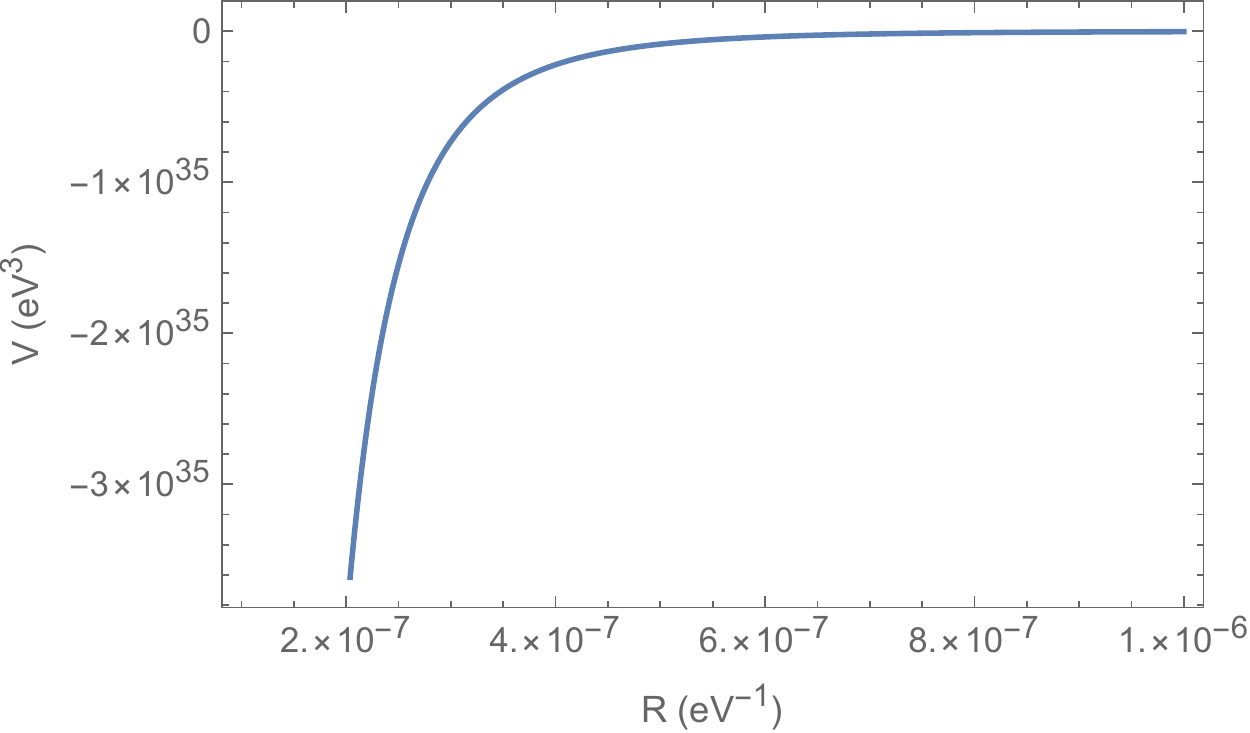}
    \caption{Radion potential at $m_3=0$ (IH). The upper panel is at $\theta = 0.196$. The lower panel is at $\theta = 0.197$. The potential becomes run-away toward $R \to 0$ for larger $\theta$.}
    \label{fig:hogehoge2}
\end{figure}

The similar numerical analysis with IH leads to the condition $ \theta < 0.0435(5)$ or $\theta>0.197(1)$ to satisfy the non-SUSY AdS conjecture (by further adjusting the mass of the neutrino). See Figs \ref{fig:hogehoge2} for the illustration of what happens near the upper bound. The constraint is severer in IH than in NH.

We have now seen that even if the lightest neutrino is Dirac, we cannot satisfy the non-SUSY AdS conjecture for all values of $\theta$. As a consequence, in order to retain the swampland conjectures, we have to make the further assumption that the allowed values of $\theta$ is restricted in the landscape. This is naturally possible when the $U(1)_{B-L}$ is broken down to its subgroup. When the $U(1)_{B-L}$ symmetry is broken down to $\mathbb{Z}_{2m}$, the allowed value of $\theta$ becomes quantized as $\theta=n/2m$ $(n=0,1,\cdots 2m-1)$.\footnote{To be more precise, if we include the quark sector, the subgroup $\mathbb{Z}_{2m}$ here actually means $\mathbb{Z}_{6m}$. Since  the fermion parity is always conserved, the order of the subgroup must be even.}

In NH, when the $U(1)_{B-L}$ is broken down to $\mathbb{Z}_2$, $\mathbb{Z}_4$, $\mathbb{Z}_8$ or $\mathbb{Z}_{10}$, we can exclude the disallowed region of $(\frac{1}{8}<)0.131<\theta <0.197 (<\frac{1}{5})$. Furthermore, for each discrete symmetry, we have to assure that the three-dimensional AdS vacuum is not realized by adjusting the mass of the lightest neutrino. For $\mathbb{Z}_2$ and $\mathbb{Z}_4$ in NH, it must be $m_1 \lesssim 8.3 \ \mathrm{meV}$ (from the constraint at $\theta = 0$); for $\mathbb{Z}_8$, it must be $m_1 \lesssim 4.5 \ \mathrm{meV}$ (from the constraint at $\theta = \frac{1}{8}$; see Fig \ref{fig:hogehoge}); for $\mathbb{Z}_{10}$, it must be $m_1 \lesssim 6.3  \ \mathrm{meV}$  (from the constraint at $\theta = \frac{1}{10}$). 

Similarly in IH, when the $U(1)_{B-L}$ is broken down to $\mathbb{Z}_2$ or $\mathbb{Z}_4$, we can exclude the disallowed region of $0.0435<\theta<0.197$. The constraint on the mass is  $m_3 \lesssim 2.8  \  \mathrm{meV}$ in both cases (from the constraint at $\theta = 0$). Note that both in NH and IH, $\mathbb{Z}_2$ is possible, but the $\mathbb{Z}_2$ symmetry allows Majorana mass, so imposing only $\mathbb{Z}_2$ while keeping the Dirac mass (without the Majorana mass) is not attractive from the naturalness viewpoint.

The $U(1)_{B-L}$ symmetry in the standard model stands out as a possible anomaly-free exact global symmetry. It could be an exact symmetry of the quantum gravity. On the other hand, the best understood swampland conjecture is the absence of the   (continuous) global symmetry \cite{Banks:2010zn}\cite{Harlow:2018jwu}\cite{Harlow:2018tng}\cite{Cordova:2022rer}, and it implies that the $U(1)_{B-L}$ cannot be an exact symmetry. Our results are in compatible with the   latter claim: the validity of the AdS swampland conjecture or the AdS distance conjecture crucially relies on the   (quantum) gravitational breaking of the $U(1)_{B-L}$ symmetry.

We have also studied the AdS distance conjecture with the twisted boundary conditions. Our study is in parallel with the one in \cite{Gonzalo:2021zsp} with the additional twist. With the homogeneous scanning \eqref{homoscan}, we have obtained the same constraints on the mass of the Dirac neutrino and the symmetry breaking as studied so far in this section. In the non-homogeneous scanning, which only changes the lightest neutrino mass, we also get the same constraint on the mass of the Dirac neutrino.\footnote{We would like to additionally note that the Majorana neutrino is consistent with the AdS distance conjecture  in the non-homogeneous scanning (but not in the homogeneous scanning) because it turns out that the vacuum is always AdS and we cannot take the Minkowski limit.}

\section{Discussions}
In this paper, we have shown that the swampland conjectures applied to a circle compactification with the $U(1)$ symmetry twisting lead to further constraints. In particular, the $U(1)$ symmetry must be broken down to $\mathbb{Z}_4$,$\mathbb{Z}_8$ or $\mathbb{Z}_{10}$ in the case of normal hierarchy, and $\mathbb{Z}_4$ in the case of inverted hierarchy, providing evidence for the absence of continuous global symmetry in quantum gravity. We also predict a more constrained upper bound  of the neutrino mass for each case.

We would like to end this paper with one important question to be addressed. In this paper, we have assumed that the only moduli to be stabilized is the radion, but this assumption implies that we have neglected the Wilson loops associated with the gauge fields in the standard model. As discussed in the seminal paper \cite{Hamada:2017yji}, it is typically the case that the non-trivial  Wilson loop leads to another vacuum with the lower energy that can be three-dimensional AdS space. While this extra vacuum cannot weaken the constraint we have obtained in this paper, it might give further constraints. In the worst scenario, the compactified standard model might not be compatible with the swampland conjectures at all (irrespective of neutrino mass). It is worthwhile studying further on this point.\footnote{One may remove the Wilson loop degrees of freedom by considering the orbifold compactification rather than the circle compactification \cite{Gonzalo:2018tpb}. However, this is not an excuse because the philosophy of the swampland conjecture states that the constraints should apply not only to one particular solution but also every solution.}

\section*{Acknowledgements}
Yu Nakayama would like to thank G.~Shiu for the correspondence and discussions.
This work by YN is in part supported by JSPS KAKENHI Grant Number 21K03581.

\end{document}